\documentclass[a4paper,twocolumn,11pt,accepted=2025-06-04]{quantumarticle}
\pdfoutput=1
\usepackage[utf8]{inputenc}
\usepackage[english]{babel}
\usepackage[T1]{fontenc}
\usepackage{amsmath}
\usepackage{hyperref}

\usepackage{tikz}
\usepackage{lipsum}

\usepackage{graphicx}
\usepackage{dcolumn}
\usepackage{bm}
\usepackage[colorlinks=true]{hyperref}
\usepackage{amssymb, amsmath, mathtools}
\usepackage{dsfont}

\usepackage[english]{babel}
\usepackage[utf8]{inputenc}
\usepackage{blindtext}
\usepackage{graphicx}
\usepackage{physics}
\usepackage{siunitx}
\usepackage{xcolor}
\usepackage[normalem]{ulem}
\usepackage{float}
\usepackage{pgfplots}
\usepackage{dcolumn}
\usepackage{multirow}

\usepackage[numbers,sort&compress]{natbib}

\newcolumntype{d}[1]{D{.}{.}{#1}}
\newcommand\mc[1]{\multicolumn{1}{c}{#1}} 

\newcommand{\pidx}[1]{{\mbox{\tiny $(#1)$}}}

\begin{document}

\title{Accurate neural quantum states for interacting lattice bosons}

\author{Zakari Denis}
\email{Email: zakari.denis@epfl.ch}
\orcid{0000-0002-4355-8800}
\affiliation{Institute of Physics, \'Ecole Polytechnique F\'ed\'erale de Lausanne (EPFL), CH-1015 Lausanne, Switzerland}
\affiliation{Center for Quantum Science and Engineering, \'Ecole Polytechnique F\'ed\'erale de Lausanne (EPFL), CH-1015 Lausanne, Switzerland}

\author{Giuseppe Carleo}
\email{Email: giuseppe.carleo@epfl.ch}
\orcid{0000-0002-8887-4356}
\affiliation{Institute of Physics, \'Ecole Polytechnique F\'ed\'erale de Lausanne (EPFL), CH-1015 Lausanne, Switzerland}
\affiliation{Center for Quantum Science and Engineering, \'Ecole Polytechnique F\'ed\'erale de Lausanne (EPFL), CH-1015 Lausanne, Switzerland}

\maketitle

\begin{abstract}
  In recent years, neural quantum states have emerged as a powerful variational approach, achieving state-of-the-art accuracy when representing the ground-state wave function of a great variety of quantum many-body systems, including spin lattices, interacting fermions or continuous-variable systems. However, accurate neural representations of the ground state of interacting bosons on a lattice have remained elusive. We introduce a neural backflow Jastrow Ansatz, in which occupation factors are dressed with translationally equivariant many-body features generated by a deep neural network. We show that this neural quantum state is able to faithfully represent the ground state of the 2D Bose-Hubbard Hamiltonian across all values of the interaction strength. We scale our simulations to lattices of dimension up to $20{\times}20$ while achieving the best variational energies reported for this model. This enables us to investigate the scaling of the entanglement entropy across the superfluid-to-Mott quantum phase transition, a quantity hard to extract with non-variational approaches.
\end{abstract}

\section{Introduction}

Lattice bosonic systems are important in condensed matter because of their unique properties. They are not subject to the exclusion principle, which allows for the appearance of smooth phases of matter in diluted settings. These phases can often be described by classical fields, which simplifies the understanding of many-body phenomena, including the study of $U(1)$ symmetry breaking, superfluidity, and boson condensation. However, simulating these systems beyond mean-field approximations is a challenge. The quantum statistics of bosons result in an exponential growth of the Hilbert space dimension with system size and particle number, complicating simulations. In driven systems, an infinite-dimensional Hilbert space might be needed for a single bosonic mode, limiting the feasibility of exact diagonalization.

Several methods have been proposed to efficiently simulate interacting lattice bosons at zero temperature. Among earliest approaches, path-integral techniques were proposed at finite temperature~\cite{batrouni1990,krauth1991a,scalettar1991,batrouni1992}. In particular, quantum Monte Carlo (QMC) simulations using the worm algorithm provided remarkably accurate predictions extrapolated to zero temperature~\cite{capogrosso-sansone2007,capogrosso-sansone2008}. At zero temperature, Green-function Monte Carlo particularly stands out~\cite{trivedi1990,calandrabuonaura1998,becca2017}, while the reptation quantum Monte Carlo algorithm~\cite{baroni1999} was extended to lattice systems~\cite{carleo2010}. However, these techniques do not give direct access to the wave function and restrict the structure of operators which may be efficiently evaluated in the chosen computational basis. Furthermore, they suffer from the sign problem~\cite{troyer2005}, even for bosonic systems in cases of geometric frustration~\cite{henelius2000,hauke2010,pan2024}. This is more generally the case with non-stoquastic Hamiltonians, such as those of bosonic lattices subjected to artificial magnetic fluxes, which have attracted consistent theoretical interest~\cite{oktel2007,moller2010,huber2011,dhar2012,dhar2013,tokuno2014,kolley2015,he2015,romen2018,zeng2020,song2020,buser2020,halati2023} and are of current experimental relevance~\cite{jaksch2003,gerbier2010,aidelsburger2013,miyake2013,struck2013,atala2014,goldman2014,sterdyniak2015,barbiero2023}.

Variational methods are more flexible in this regard. Variational Monte Carlo (VMC) using a modified Jastrow Ansatz~\cite{capello2005,capello2007,capello2008} reproduced all of the features of the quantum phase transition in all relevant spatial dimensions, though only qualitatively. Tensor networks have been used, in particular, to probe ground-state properties of lattice bosons~\cite{alba2013,rams2018,weerda2024}. However, this is restricted by entanglement and typically requires to introduce truncations in the local occupation factors. The generic area-law entanglement scaling of lattice bosons~\cite{eisert2010} and the growth of entanglement upon real-time propagation restrict the efficiency of matrix-product variational representations mainly to the ground state of one-dimensional geometries, such as open chains~\cite{pai1996,kuhner1998,rapsch1999,kollath2007,lauchli2008} or small-perimeter cylinders~\cite{alba2013}, and short time evolutions.

In recent years, neural quantum states~\cite{carleo2017} have emerged as a powerful variational method, consistently demonstrating remarkable accuracy in representing the ground-state wave function of a wide range of nontrivial Hamiltonians. In addition to spin problems, where it stands as the state-of-the-art method for frustrated lattices~\cite{nomura2021,chen2024,rende2024}, properly tailored networks have demonstrated their effectiveness in addressing problems involving other kinds of degrees of freedom, including bosonic~\cite{pescia2022}, fermionic~\cite{hermann2020,pfau2020,pescia2024} and beyond~\cite{luo2021,medvidovic2023}. Despite these successes, highly accurate neural-network representations of the ground state of lattice bosonic systems have remained elusive. Several works~\cite{saito2017,mcbrian2019,vargas-calderon2020,zhu2023} have recently contributed to these ongoing efforts, where the most notable advances~\cite{choo2018,pei2024} have been achieved by incorporating some of the physical structure of the system in the variational Ansatz.

In this work, we introduce a new bosonic neural Ansatz based on the concept of backflow~\cite{feynman1956}, originally formulated in continuous space~\cite{lee1981,kwon1993,kwon1998,holzmann2003} and later extended to lattices~\cite{tocchio2008,tocchio2011}. Neural-network parametrizations of the backflow transformation~\cite{ruggeri2018,luo2019} have driven the most recent advances in fermionic neural quantum states~\cite{hermann2020,pfau2020,pescia2024}.
We provide analytical support to motivate its structure and benchmark it on the celebrated Bose-Hubbard model in two dimensions. We assess its remarkable accuracy and scale our simulations to lattices with up to $20{\times}20$ sites without local truncation, allowing us to perform a finite-size scaling analysis around the critical point. We further investigate the entanglement entropy of the system across the phase transition, a matter of theoretical~\cite{alba2013,metlitski2015,frerot2016} and experimental~\cite{islam2015} interest.

The paper is structured as follows: the Bose-Hubbard physics is briefly described in Sec.~\ref{sec:2}; in Sec.~\ref{sec:3}, we introduce the neural backflow-Jastrow architecture and formally derive the structure of some of the correlations it can capture; we then apply it to simulate the ground state of the Bose-Hubbard Hamiltonian in Sec.~\ref{sec:4}, before finally concluding in Sec.~\ref{sec:5}.

\section{Bose Hubbard\label{sec:2}}

In this work, we focus on the paradigmatic Bose-Hubbard model~\cite{gersch1963}, as given by the Hamiltonian

\begin{equation}
    \hat{H} = -J\sum_{\langle i,j\rangle}(\hat{a}_i^\dagger\hat{a}_j^{\mathstrut} + \hat{a}_j^\dagger\hat{a}_i^{\mathstrut}) + \frac{U}{2}\sum_i\hat{a}_i^\dagger\hat{a}_i^\dagger\hat{a}_i^{\mathstrut}\hat{a}_i^{\mathstrut}.
    \label{eq:bose-hubbard}
\end{equation}
Here, $\hat{a}_i$ denotes the annihilation operator at site $i$ of the lattice under consideration, satisfying the usual bosonic canonical commutation relations, $J$ is the rate of the hopping of bosons between nearest-neighbor sites, and $U$ quantifies the strength of on-site repulsive interactions. This system can be realized in a number of platforms, notably with ultra-cold atoms loaded in an optical lattice~\cite{jaksch1998,greiner2002,spielman2007,spielman2008,bloch2008,endres2012,islam2015}, or Josephson-junction arrays~\cite{vanotterlo1993,bruder2005}. In what follows, we shall restrict ourselves to $L{\times}L$ two-dimensional square lattices at fixed number of particles $N$ and unit density $\bar{n} := N/L^2 = 1$.

This system displays a second-order phase transition resulting from the competition between the kinetic energy and the potential energy. At infinite interaction strength and integer density, the wavefunction is a product of Fock states: the so-called Mott phase. As the ratio $J/U$ increases, the $U(1)$ symmetry under the global transformation $\hat{a}_i\mapsto \hat{a}_i e^{i\alpha}$ spontaneously breaks, and the ground state becomes extensively degenerate. The bosonic population spontaneously condensates into the zero-momentum mode, as revealed by a finite value of the condensate fraction:
\begin{equation}
    \rho_0 := \langle\hat{n}(\vb*{k}=0)\rangle / L^2 \equiv \sum_{ij}\langle\hat{a}_i^\dagger\hat{a}_j^{\mathstrut}\rangle/L^4,
\end{equation}
which here plays the role of order parameter~\cite{yang1962}. In 2D, this phase transition is known to belong to the 3D XY-model universality class~\cite{fisher1989,batrouni1990,batrouni1992,krauth1991,capogrosso-sansone2008}. The critical value of the control parameter, $J_c/U = 0.05974(3)$ [$U_c/J = 16.74(1)$], was first precisely determined by a high-order strong coupling expansion~\cite{freericks1996,elstner1999} and later refined by means of QMC calculations extrapolated to vanishing temperature~\cite{capogrosso-sansone2008}.

\section{Variational model\label{sec:3}}

The simplest many-body variational Ansatz capable of qualitatively describing the singular depletion of the condensate while also reproducing the proper behavior of the spatial correlations is the two-body Jastrow wavefunction~\cite{bijl1940,jastrow1955}
\begin{equation}
    \ket{\psi_J} = e^{\sum_{ij}\hat{n}_{i}W_{ij} \hat{n}_{j}}\ket{\psi_0},
    \label{eq:jastrow}
\end{equation}
where $\ket{\psi_0} = N^{-1/2}(\sum_i\hat{a}_i^\dagger)^N\ket{\vb*{0}}$ is the wavefunction of the ideal condensate, and $\vb{W}$ is a $L^2\times L^2$ matrix of variational parameters. This can be equivalently written as
\begin{equation}
    \ln\psi_J(\vb*{n}) = \sum_{ij}n_{i}W_{ij} n_{j} + \ln \psi_0(\vb*{n}),
\end{equation}
where $\vb*{n} := (n_1, \ldots, n_{L^2})$ and $\psi(\vb*{n}) \equiv \braket{\vb*{n}}{\psi}$.

In Refs.~\cite{capello2005,capello2007,capello2008}, authors showed that the variational accuracy could be greatly enhanced without losing the physical structure and intuition of the two-body Jastrow Ansatz. This is done by adding the following many-body Gutzwiller projector:
\begin{equation}
    \ket{\psi_\mathrm{MBJ}} = e^{g_\mathrm{MB}(\hat{\Pi}_h + \hat{\Pi}_d)}\ket{\psi_\mathrm{J}},
    \label{eq:mbj}
\end{equation}
where $g_\mathrm{MB}$ is a real variational parameter and the projectors $\hat{\Pi}_h = \sum_i \hat{h}_i\bigotimes_{j\in N(i)}(\hat{\mathds{1}}-\hat{d})_j$ and $\hat{\Pi}_d = \sum_i \hat{d}_i\bigotimes_{j\in N(i)}(\hat{\mathds{1}}-\hat{h})_j$, with $\hat{h} = \ketbra{0}$ and $\hat{d} = \ketbra{2}$, and $N(i)$ denoting the neighborhood of $i$ on the lattice, measure the amount of isolated holons and doublons, respectively. This many-body term serves as a penalty for localized density inhomogeneities, spatially binding holon-doublon pairs, and thereby favoring the onset of a homogeneous Mott insulating phase.

This variational Ansatz, although simple, is able to qualitatively capture the physics of the transition, including the behavior of the correlations and the structure factor~\cite{capello2005,capello2007,capello2008}. However, the quantitative agreement is not excellent, with an estimated ground-state energy for this Ansatz affected by a relative error of the order of $10\%$ with respect to the exact reference and the transition point shifted by $25\%$ to around $U_c/J \sim 21$.

In what follows, we introduce a flexible generalization of the above construction based on the concept of neural backflow transformation~\cite{luo2019}.

\subsection{Many-body Jastrow}

The most straightforward improvement over the Ansatz of Eq.~\eqref{eq:mbj} is to lift the Jastrow term to a $M$-body term. Let us show that the hierarchy of correlations that can be accounted for by a deep convolutional neural network (CNN) encompasses that of such a many-body Jastrow term.

Let us consider a network of depth $D$ built upon the following single-channel convolutional layer with skip connections:
\begin{align}
    h_i^\pidx{n}(\vb*{n}) &= n_i + \tilde{h}_i^\pidx{n}(\vb*{n}),\\
    \tilde{h}_i^\pidx{n}(\vb*{n}) &= f\bigl({\textstyle\sum_j} w_{\vb*{r}_{ij}}^\pidx{n}h_j^\pidx{n{-}1}(\vb*{n}) + b^\pidx{n}\bigr),
\end{align}
with $W_{\vb*{r}} = w_{\vb*{r}} = 0$ for any $\vb*{r}$ such that $\lvert\vb*{r}\rvert > d_F$, where $d_F$ corresponds to the dimension of the convolutional filter. Upon assuming for simplicity that the activation function $f$ belongs to $C^2$ and possesses a non-vanishing second-order derivative, one has the following recursion relation\footnote{Here, the Lagrange form of the first-order Taylor expansion of $f$ around the bias $b$ was used: $\tilde{h}^{\pidx{n}}_i = f(b) + {\sum_j} \bigl[f'(b) w_{\vb*{r}_{ij}}\bigr]h^{\pidx{n{-}1}}_j + \frac{1}{2}{\sum_{jk}} \bigl[f''(b_\star)w_{\vb*{r}_{ij}}w_{\vb*{r}_{ik}}\bigr] h^{\pidx{n{-}1}}_j h^{\pidx{n{-}1}}_k$. Interestingly, the mean-value theorem, at the core of the Lagrange remainder, was also used to show that the ground state of a system of interacting particles in continuous space is exactly captured by a general backflow transformation~\cite{pescia2024}.}:
\begin{widetext}
\begin{gather}
    \sum_i h_i^\pidx{n}(\vb*{n}) = \ldots + \frac{1}{2}\sum_{i_0 i_1} n_{i_0} W_{i_0 i_1}^\pidx{n} h_{i_1}^\pidx{n{-}1}(\vb*{n}) = \ldots + \sum_{n=1}^D\frac{1}{2^n}\sum_{i_0 \cdots i_n}n_{i_0}W^\pidx{n}_{i_0 i_1}n_{i_1}W^\pidx{n{-}1}_{i_1 i_2}\ldots W^\pidx{1}_{i_{n{-}1} i_n}n_{i_n},
\end{gather}
\end{widetext}
that is a linear combination of $n$-body Jastrow terms with $n\leq D+1$, with symmetric two-body translationally equivariant Jastrow weights given by
\begin{equation}
    W_{ij}^\pidx{n} \equiv W_{ji}^\pidx{n} \propto \sum_k w_{\vb*{r}_{ki}}^\pidx{n}w_{\vb*{r}_{kj}}^\pidx{n}.\\
\end{equation}
Trivially, the two-body Jastrow weights can be made independent for each $n$-body term at the cost of at most $M$ channels in the CNN.

An analogous result is obtained, replacing skip connections by residual connections, as will be later introduced. Therefore, range-$d_F$ $M$-body Jastrow factors can be generated with a residual neural-network (ResNet) Ansatz with depth $M-1$ and filters of width $d_F$.

\subsection{Generalized many-body Gutzwiller projector}

In Eq.~\eqref{eq:mbj}, the many-body term acts as a partial projection of the wave function onto a manifold with specific values of the local quantum numbers within some finite-size patch. The choice of the specific values of these quantum numbers is set from \textit{a priori} knowledge about the underlying physics. We may loosen this inductive bias at the cost of a larger number of variational parameters.

Let us consider the following many-body Gutzwiller Ansatz:
\begin{equation}
    \ket{\psi_\mathrm{MB}} = e^{g_\mathrm{MB}\sum_i\bigotimes_{\vb*{v}\in \mathcal{V}} \ketbra{\smash{x_{\vb*{v}}}}_{T_{\vb*{v}}(i)}}\ket{\psi_0},
\end{equation}
where $g_\mathrm{MB}$ is a free variational parameter, $\mathcal{V}$ denotes a set of relative positions with respect to the $i$th site, thereby defining a \textit{patch}, and $\mathcal{X} = \lbrace x_{\vb*{v}}, \vb*{v} \in \mathcal{V}\rbrace$ some corresponding fixed local configurations given \textit{a priori}. Here, $T_{\vb*{v}}(i)$ denotes the site corresponding to $i$ translated on the lattice by $\vb*{v}$. This Ansatz may be evaluated as
\begin{align}
    \ln\psi_\mathrm{MB}(\vb*{n}) =& \ln \Pi_\mathrm{MB}(\vb*{n}) + \ln\psi_0(\vb*{n}),\\
    \ln \Pi_\mathrm{MB}(\vb*{n}) :=& g_\mathrm{MB} \sum_i \prod_{\vb*{v}\in\mathcal{V}}\mathds{1}[n_{T_{\vb*{v}}(i)} = x_{\vb*{v}}];\label{eq:lnmb}
\end{align}
where $\mathds{1}$ denotes the indicator function. The diagonal many-body projector of Eq.~\eqref{eq:lnmb} may be represented by a CNN. Indeed, one has that
\begin{gather}
    \ln \Pi_\mathrm{MB}(\vb*{n}) = \sum_{i}h_{i,1}^\pidx{2}(\vb*{n}),\\
    h_{i,1}^\pidx{2}(\vb*{n}) = \mathrm{ReLU}\Bigl(-2g_\mathrm{MB}\Delta x\sum_{\mu=1}^2 h_{i,\mu}^\pidx{1}(\vb*{n})+g_\mathrm{MB}\Bigr),\\
    h_{i,\mu}^\pidx{1}(\vb*{n}) = \mathrm{ReLU}\Bigl(\sum_{\vb*{v}\in\mathcal{V}} \frac{(-1)^\mu}{x_{T_{\vb*{v}}(i)}}n_{T_{\vb*{v}}(i)} - (-1)^{\mu}\Bigr),
\end{gather}
where $h_{i,\mu}^\pidx{n}$ denotes the $\mu$th channel of the $i$th output of the $n$th layer of the CNN, and $\Delta x > \max_{x\in\mathcal{X}}(x)$\footnote{The case where the value $0$ belongs to $\mathcal{X}$ can be simply worked out by a fixed initial linear transformation of all inputs as $n_i \mapsto n_i + \xi$, $\forall \xi \in \mathbb{R}_+^\star$. In particular, Eq.~\eqref{eq:mbj} is recovered using local fluctuations $n_i - \bar{n}$ as an input and identifying holes (doublons) with the value $x = -1$ ($x=+1$)}.
This corresponds to a two-layer CNN with parameters:
\begin{align}
    K^\pidx{1}_{\vb*{v},\mu,1} &= (-1)^\mu/x_{T_{\vb*{v}}(i)} \in \mathbb{R}^{\lvert\mathcal{V}\rvert\times 2\times 1},\nonumber\\
    b_\mu^\pidx{1} &= (-1)^{\mu+1}\in\mathbb{R}^2,\nonumber\\
    K^\pidx{2}_{\vb*{0},1,\mu'} &= -2g_\mathrm{MB}\Delta x \in \mathbb{R}^{1\times1\times2},\nonumber\\
    b_1^\pidx{2} &= g_\mathrm{MB}\in\mathbb{R}.
    \label{eq:params}
\end{align}

Therefore, a many-body Gutzwiller Ansatz with $K$ many-body projectors may be represented by a two-layer CNN with a filter geometry matching $\mathcal{V}$ and $K$ channels. Relaxing the conditions in Eq.~\eqref{eq:params} generalizes the projection to the optimum quantum numbers. In App.~\ref{app:a}, we show how this architecture can also account for an attractive confining potential for holons and doublons close apart, effectively stabilizing the Mott phase.

\begin{figure}
    \centering
    \includegraphics[width=.85\linewidth]{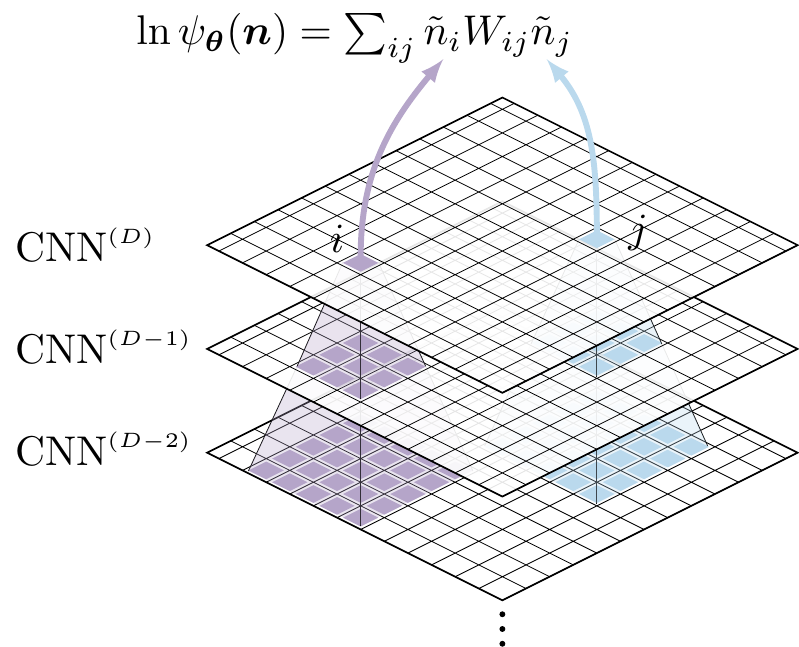}
    \caption{Schematic representation of the neural backflow-Jastrow architecture of Eq.~\eqref{eq:bfj}. The backflow transformation dresses the bare local quantum numbers entering the Jastrow term with equivariant many-body features. These features bear a dependence on the configuration of all sites within the network's receptive field. The latter is represented by the shaded region for an exemplary deep convolutional neural network with square filters of width $d_F = 3$.}
    \label{fig:1}
\end{figure}

\subsection{Neural backflow Jastrow Ansatz}

This motivates the use of a NQS Ansatz of the form~\cite{choo2018}
\begin{equation}
    \ln\psi_{\vb*{\theta}}(\vb*{n}) = \sum_{i,\mu} h_{i,\mu}^\pidx{D}(\vb*{n}) + \ln\psi_0(\vb*{n}),
\end{equation}
where $\vb*{\theta}$ denotes the set of all variational parameters, and $h_{i,\mu}$ is the channel $\mu$ of site $i$ of a depth-$D$ residual network with stride equal to one.

Provided the number of channels is large enough, such a network should be able to encompass both the $M$-body Jastrow term and the many-body Gutzwiller projector. However, the range of the Jastrow terms in such a parametrization is limited by the width of the filters $d_F$. This is in contrast with the initial Jastrow Ansatz of Eq.~\eqref{eq:jastrow} which could capture the crucial long-range correlations responsible of the singular behavior of the density structure factor at zero momentum~\cite{capello2007}.

To circumvent this, we instead introduce the many-body features of the ResNet as an equivariant backflow transformation of a translationally invariant two-body Jastrow with maximum range. This leads to the variational Ansatz that we shall consider throughout this article:
\begin{equation}
    \ln \psi_{\vb*{\theta}}(\vb*{n}) = \sum_{i,j = 1}^{L^d}\tilde{n}_{i}W_{d_{ij}} \tilde{n}_{j},
    \label{eq:bfj}
\end{equation}
where the weights only depend on the $L^1$ minimum-image distance $d_{ij}$ between all pairs $(i,j)$. We also dropped the original mean-field prior $\psi_0$ which proves useful only for small system sizes. Fed to this Jastrow factor are translation-equivariant many-body features obtained according to the following backflow transformation of the original local occupation factors:
\begin{equation}
    \tilde{n}_{i} = n_{i} + \sum_{\mu=1}^{\alpha_D} a_\mu h_{i,\mu}^\pidx{D}.
    \label{eq:backflow}
\end{equation}
Here, $h_{i,\mu}^\pidx{D}$ are translation-equivariant features extracted from the original occupations through a residual neural network of depth $D$ with $\alpha_D$ channels, and the mixing weights $a_\mu$ both combine all channels and control the magnitude of the backflow transformation. The former are obtained as $h_{i,\mu}^\pidx{2\ell} = \tilde{h}_{i,\mu}^\pidx{2\ell{-}1}$ and $h_{i,\mu}^\pidx{2\ell {+} 1} = \mathrm{LayerNorm}(h_{i,\mu}^\pidx{2\ell {-} 1} + \tilde{h}_{i,\mu}^\pidx{2\ell})$, with
\begin{equation}
    \tilde{h}_{i,\mu}^\pidx{\ell} = \sigma\Biggl(\sum_{\mu'=1}^{\alpha_\ell}\sum_{\substack{\vb*{v}\,:\, \mathrlap{v_\infty\leq d_K}}}K_{\vb*{v},\mu\mu'}^\pidx{\ell}h_{T_{\vb*{v}}(i),\mu'}^\pidx{\ell{-}1} + b_\mu^\pidx{\ell}\Biggr),
\end{equation}
where, $\sigma$ is an activation function ($\mathrm{GELU}$), $\alpha_\ell$ the number of channels of the $\ell$th layer, $d_K$ the kernel size, $T_{\vb*{v}}(i)$ corresponds to the site $i$ translated by $\vb*{v}$, and $(K_{\vb*{v},\mu\mu'}^\pidx{\ell}, b_\mu^\pidx{\ell})$ the filters and bias of the $\ell$th layer. To further improve stability, the input occupations are rescaled as $n_i \mapsto n_i/\bar{n} - 1$, $\forall i$, with $\bar{n} = N_p / L^d$. The architecture of the model is illustrated in Fig.~\ref{fig:1}.

Let us note that this construction, in presence of a bias term in the last layer of the backflow transformation, is at least as expressive as the bare ResNet. Furthermore it allows one to perform VMC more efficiently by initializing the weights of the Jastrow with those obtained by first optimizing a bare two-body Jastrow Ansatz of the form of Eq.~\eqref{eq:jastrow}, and then optimizing the entire network altogether. Indeed, provided the backflow weights are initialized such that they initially act perturbatively, this allows to start the optimization closer from the ground state and thus with a much lower variance on all Monte Carlo estimates involved in the process.

As detailed in App.~\ref{app:b}, splitting the layout of the wavefunction into a simple mean-field-like long-range behavior and a complex short-range structure is a convenient design choice performance-wise, especially for gapped Hamiltonians whose ground states exhibit some finite-range quantum correlations.

Very recent variational Ansätze based upon the visual-transformer architecture~\cite{viteritti2023,rende2024,viteritti2025} adopt a reminiscent structure, rooted in representation learning~\cite{bengio2013}, where a strong scission of the model into a deep embedding and a shallow final correlator proves optimal.

\section{Numerical results\label{sec:4}}

In what follows, we optimize the parameters $\vb*{\theta}$ of the model defined by Eq.~\eqref{eq:bfj} by minimizing the variational expectation value of the energy
\begin{equation}
    E_{\vb*{\theta}} = \frac{\bra{\psi_{\vb*{\theta}}}\hat{H}\ket{\psi_{\vb*{\theta}}}}{\braket{\psi_{\vb*{\theta}}}},
\end{equation}
with $\hat{H}$ as given by Eq.~\eqref{eq:bose-hubbard}. This is achieved with guarantees of exponential convergence via the stochastic-reconfiguration prescription~\cite{sorella2005}, where the parameters are evolved in time according to
\begin{equation}
    \dot{\vb*{\theta}} = -\vb{S}^{-1}\vb*{F},
\end{equation}
where $\vb{S}$ and $\vb*{F}$ respectively denote the quantum geometric tensor~\cite{stokes2020} and the vector of forces, as given by
\begin{gather}
    S_{kk'} = \mathbb{E}\bigl[O_k^\star(\vb*{n}) O_{k'}(\vb*{n})\bigr] - \mathbb{E}\bigl[O_k^\star(\vb*{n})] \mathbb{E}\bigl[O_{k'}(\vb*{n})\bigr],\\
    F_k = \mathbb{E}\bigl[O_k^\star(\vb*{n}) E_{\mathrm{loc}}(\vb*{n})\bigr] - \mathbb{E}\bigl[ O_k^\star(\vb*{n})\bigr] \mathbb{E}\bigl[E_{\mathrm{loc}}(\vb*{n})\bigr],
\end{gather}
with $O_{k}(\vb*{n}) := \partial_{\theta_k}\ln\psi_{\vb*{\theta}}(\vb*{n})$ the log-derivative of the variational Ansatz and $E_\mathrm{loc}(\vb*{n}) := \bra{\vb*{n}}\hat{H}\ket{\psi_{\vb*{\theta}}}/\braket{\vb*{n}}{\psi_{\vb*{\theta}}}$ the so-called \textit{local energy}. In the above, expectation values are implicitly taken with respect to the Born distribution: $\mathbb{E}[f(\vb*{n})] := \sum_{\vb*{n}} f(\vb*{n}) \lvert\psi_{\vb*{\theta}}(\vb*{n})\rvert^2/\braket{\psi_{\vb*{\theta}}}$.

Importantly, all of these quantities can be efficiently estimated by drawing samples from this probability density function thanks to, e.g., Markov chain Monte Carlo and the Metropolis-Hastings algorithm. More details about the variational optimization are given in App.~\ref{app:c}.

In all following simulations, ResNet-based backflow transformations with square filters of width $d_F=3$ and $\alpha = 12$ channels per layer were used. A depth of $D=8$ was used for $20{\times}20$ lattices and $D=6$ for the rest. All models were optimized through NetKet 3.0~\cite{vicentini2022} on a single GPU.

\addtolength{\tabcolsep}{-1pt}
\begin{table}[t]
\centering
\resizebox{\linewidth}{!}{
\begin{tabular}{crccl}
Size & \mc{$U/J$} & Method & Parameters & \mc{$E_{\vb*{\theta}} / JL^{2}$}\\
\hline
\multirow{4}{*}{$8{\times}8$} & \multirow{2}{*}{$8.5$} & GFMC~\cite{gfmc} & --- & $-1.544669(8)$\\
 & & NBFJ & $\num{6681}$ & $-1.544668(3)$\\
 & \multirow{2}{*}{$17.0$} & GFMC~\cite{gfmc} & --- & $-0.5375(4)$\\
 & & NBFJ & $\num{6681}$ & $-0.53804(2)$\\
\hline
\multirow{4}{*}{$10{\times}10$} & \multirow{2}{*}{$16.0$} & NQS-OH~\cite{pei2024} & $\num{426}$ & $-0.5721(3)$\\
 & & NBFJ & $\num{6683}$ & $-0.58972(3)$\\
 & \multirow{2}{*}{$20.0$} & NQS-B~\cite{pei2024} & $\num{206}$ & $-0.4068(9)$\\
 & & NBFJ & $\num{6683}$ & $-0.43329(3)$\\
\end{tabular}
}
\caption{\label{tab:1}Comparison of the variational energy obtained by the neural backflow Jastrow (NBFJ) Ansatz with depth $D=6$ and those of Green-function Monte Carlo (GFMC)~\cite{gfmc} and recent NQS simulations~\cite{pei2024}.}

\end{table}

\begin{figure}[!t]
    \centering
    \includegraphics[width=\linewidth]{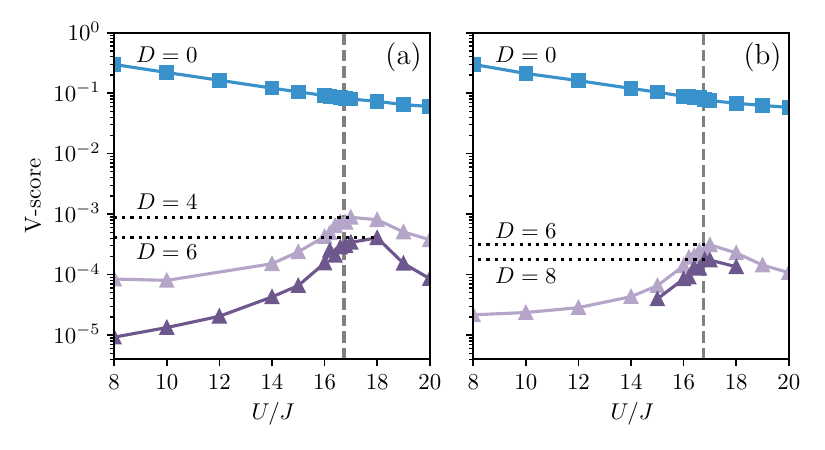}
    \caption{V-score as a function of the depth $D$ of the neural backflow transformation for lattices of size (a) $16{\times}16$ and (b) $20{\times}20$ at unit filling $\bar{n}=1$. The gray shaded region indicates the critical value of the control parameter found by QMC~\cite{capogrosso-sansone2008}.}
    \label{fig:2}
\end{figure}

\subsection{Benchmarking}

The scaling of the Hilbert-space dimension with system size is rather adverse for bosonic lattices. Therefore, exact-diagonalization reference quantities are restricted to very small systems which are not very representative of the physics at work in the thermodynamic limit.

At moderate system size, ground-state predictions may be efficiently obtained via Green-function Monte Carlo (GFMC)~\cite{trivedi1990,calandrabuonaura1998,becca2017}. These provide a valuable reference value since they are exact up to a controllable systematic bias. In Table~\ref{tab:1}, we compare the variational energies achieved by our Ansatz for a $8{\times}8$ lattice against a GFMC reference~\cite{gfmc}, showing perfect agreement. Therein, we further compare our variational energies to the best previously achieved NQS results~\cite{pei2024} for a $10{\times}10$ system, obtaining values $3$ to $6$\% lower, although with a higher number of parameters.

For larger system sizes, instead, we assess the accuracy of our wavefunction by means of the recently introduced variational score (V-score)~\cite{wu2024}, which allows to systematically compare Ansätze. For bosonic systems, this score can be computed as
\begin{equation}
    \text{V-score} = \frac{L^d \mathrm{Var}[E_{\vb*{\theta}}]}{(E_{\vb*{\theta}} - E_\mathrm{MF})},
\end{equation}
Where $L^d$ is the number of sites of the system, $\mathrm{Var}[E_{\vb*{\theta}}]$ is the variance of the variational energy, and $E_\mathrm{MF}$ is the mean-field energy. The latter is given by $E_\mathrm{MF}/L^d = \bar{n}(U\bar{n}/2 - zJ)$ for Bose Hubbard with coordination number $z$ ($z=4$ here).

This quantity is intensive. Therefore, it allows one to systematically compare the accuracy of a simulations at varying system sizes. It vanishes when the variational Ansatz exactly matches the ground state by virtue of the zero-variance property. Furthermore, it was empirically verified~\cite{wu2024} that the relative error on the variational energy of the ground state positively correlates to the V-score of the Ansatz following a universal trend. This can be used to roughly infer the error, which proves very useful whenever exact diagonalization is not available. 

A V-score in the range $10^{-4}$--$10^{-6}$ signals a relative error roughly below $10^{-5}$, indicating that the problem has been---for all practical purposes---solved variationally. This is the case, for instance, of spin models in the absence of magnetic frustration.

In Fig.~\ref{fig:2}, we show the V-score achieved by the backflow-Jastrow Ansatz for Bose Hubbard at unit filling on a periodic 2D lattice with $N=16{\times}16$ and $N=20{\times}20$ sites. The problem is clearly harder in the vicinity of the critical point, where the gap closes in the thermodynamic limit. A few ResNet layers in the backflow transformation dramatically improve the accuracy over bare Jastrow, lowering the V-score from $10^{-1}$ down to about $2{\times}10^{-4}$ for a depth of $D=8$ layers. We observe that increasing the depth $D$ of the backflow transformation consistently improves the accuracy. Interestingly, the depth required to reach a given accuracy does not seem to depend on the lattice size. Variational energies along with corresponding V-scores for all tested neural-backflow depths at $U/J = 16.8$ ($\sim U_c/J$) are provided in App.~\ref{app:d}.

\begin{figure}[!t]
    \centering
    \includegraphics[width=\linewidth]{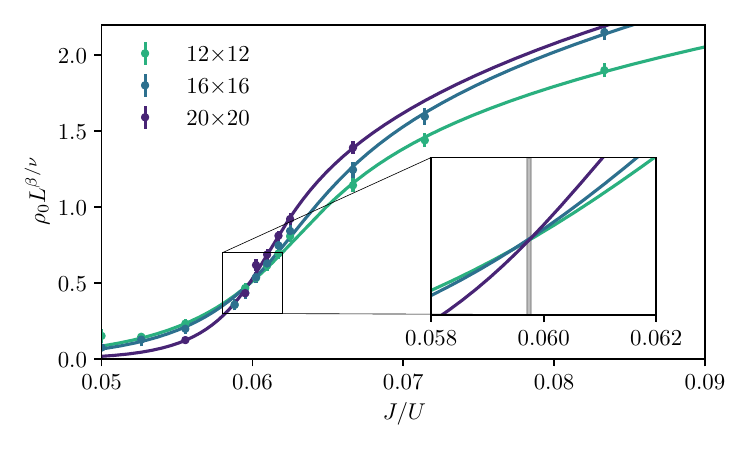}
    \caption{Rescaled condensate fraction as a function of the inverse interaction strength for increasing lattice sizes. Plain lines denote a fit of the scaling function in Eq.~\eqref{eq:scaling_function} to the variational data. The gray shaded region in the inset indicates the confidence interval for the critical value of the control parameter as found by QMC calculations~\cite{capogrosso-sansone2008}. As expected by scaling theory, all curves intersect for a same abscissa, showing good correspondence with the QMC estimation.}
    \label{fig:3}
\end{figure}

\begin{figure}[!t]
    \centering
    \includegraphics[width=\linewidth]{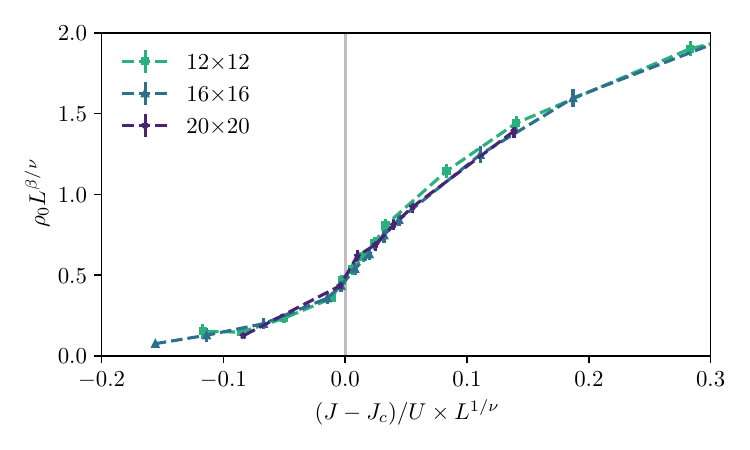}
    \caption{Data collapse of the data of Fig.~\ref{fig:3} when plotted against the rescaled centered control parameter, showing the universal behavior of observables close to the critical point.}
    \label{fig:4}
\end{figure}

\subsection{Finite-size scaling}

We can exploit the ability of our approach to scale to large system sizes to investigate the universal finite-size scaling of physical quantities of the system. According to the usual finite-size scaling argument, the order parameter $\rho_0/N$ for a lattice of linear size $L$ should depend on the control parameter $(J-J_c)/U$ as $\rho_0/N\rvert_L \sim L^{-\beta/\nu}\tilde{f}(L^{1/\nu}(J-J_c)/U)$ where $\tilde{f}$ is some scaling function. Hence, when plotting $\rho_0/N\rvert_L L^{\beta/\nu}$ against $J/U$ for various values of $L$, all curves should intersect at the critical value $J_c/U$, provided the ratio of critical exponents $\beta/\nu$ is correct.

In Fig.~\ref{fig:3}, we perform the above analysis using the critical exponents of the 3D XY universality class~\cite{campostrini2001}. Upon fitting to our variational data the fitting function
\begin{equation}
    f(J/U) = \bigl[\mathrm{softplus}_a(J/U - b)\bigr]^c,\label{eq:scaling_function}
\end{equation}
with $\mathrm{softplus}_\alpha(x) := \ln\bigl(1 + e^{\alpha x}\bigr)/\alpha$, and parameters $a$, $b\sim J_c/U$ and $c$, all curves intersect within the confidence intervals of the best available estimation of the critical value of $J/U$~\cite{capogrosso-sansone2008}. In Fig.~\ref{fig:4}, we verify that all of the variational data collapse into a universal curve upon rescaling and centering the abscissa of Fig.~\ref{fig:3}.

\begin{figure*}[!t]
    \centering
    \includegraphics[width=\linewidth]{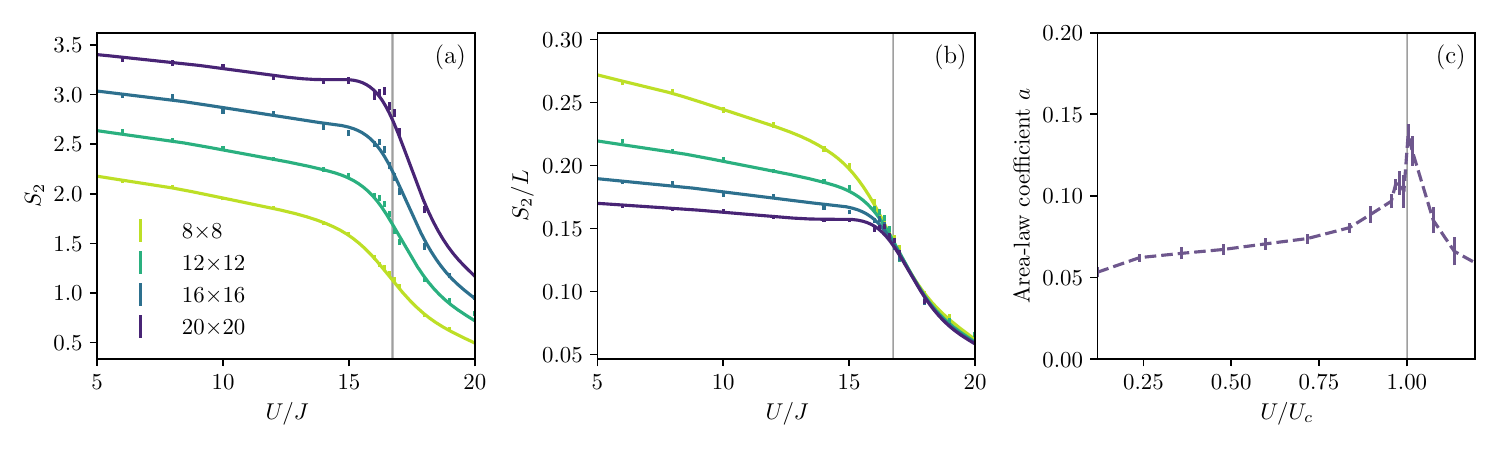}
    \caption{(a) Rényi-2 entropy $S_2(\hat{\rho}_A)$ evaluated on the reduced density matrix of the first half ($A$) of the lattice. (b) Entanglement entropy rescaled by the typical dimension of the subsystem's boundary length $\lvert\partial A\rvert \sim L$. In the Mott insulating phase, a strict area law is observed while the spontaneous breaking of the continuous symmetry induces a logarithmic correction in the superfluid phase. (c) Area-law coefficient $a$ obtained by fitting Eq.~\eqref{eq:s2-scaling} to the variational data for linear sizes $L\in\lbrace 8, 10, 12, 14, 16, 20\rbrace$. In panels (a) and (b), errors are estimated by bootstrap; a cubic spline interpolation of the $S_2$ fit is added as a guide to the eye.
    }
    \label{fig:5}
\end{figure*}

\subsection{Entanglement entropy}

One of the strengths of variational approaches is the direct access to the wavefunction, in contrast with other techniques such as QMC. This allows us to compute quantities beyond usual linear forms on the wavefunction. In particular, a quantity of interest is the entanglement entropy, which quantizes the degree of entanglement of a subsystem composed of a subset of the lattice $A$ and its complement $\bar{A}$. It is defined as the entropy of the reduced density $\hat{\rho}_A = \Tr_{\bar{A}}[\ketbra{\psi_{\vb*{\theta}}}]$. While the von Neumann entropy is typically used, the Rényi-2 entanglement entropy, as given by
\begin{equation}
    S_2(\hat{\rho}_A) = -\ln\bigl(\Tr\hat{\rho}_A^2\bigr),
\end{equation}
can be easily estimated via Monte Carlo sampling of two independent replicas through the estimator~\cite{hastings2010,torlai2018,zhao2022}
\begin{equation}
    S_2(\hat{\rho}_A) = -\ln\mathbb{E}_{\vb*{n}, \vb*{n}'} \left[ \frac{\psi_{\vb*{\theta}}(\vb*{n}'_A,\vb*{n}_{\bar{A}})\psi_{\vb*{\theta}}(\vb*{n}_A,\vb*{n}'_{\bar{A}})}{\psi_{\vb*{\theta}}(\vb*{n}_A,\vb*{n}_{\bar{A}})\psi_{\vb*{\theta}}(\vb*{n}'_A,\vb*{n}'_{\bar{A}})} \right].
    \label{eq:renyi2-estimator}
\end{equation}
In this estimator, samples $\vb*{n}$ and $\vb*{n}'$ are first drawn from two independent Markov chains and split into configurations of either complementary subspace $\vb*{n}^\pidx{\prime} = (\vb*{n}^\pidx{\prime}_A,\vb*{n}^\pidx{\prime}_{\bar{A}})$. In the numerator, the wavefunction is evaluated on configurations generated by partially swapping the configurations of the replicas according to $\vb*{n}_A \leftrightarrow \vb*{n}'_A$. Importantly, when evaluating the above estimator, we enforce that $\psi_{\vb*{\theta}}(\vb*{n}) = 0$ whenever $\sum_i n_i \neq N$. In what follows, we shall consider the first half of the lattice, of dimension $L\times L/2$ and boundary length of $\lvert\partial A\rvert = 2L$, as the subsystem $A$.

Such a system is expected to exhibit an area-law scaling of the entanglement entropy. However, at finite size, the degeneracy of the ground state in the superfluid is slightly lifted. This \textit{tower-of-states} mechanism~\cite{metlitski2015} induces a correction in the superfluid phase scaling as the logarithm of the size of the boundary, resulting into the law~\cite{alba2013}
\begin{equation}
    S_2(\hat{\rho}_A) = a L + b\mathds{1}[U>U_c] \ln L + c + O(1/L),
    \label{eq:s2-scaling}
\end{equation}
where $b = 1/2$ is a universal coefficient related to the number of Goldstone modes ($1$ for the 2D Bose-Hubbard and 3D XY models).

In Fig.~\ref{fig:5}(a), we show the Rényi-2 entropy $S_2(\hat{\rho}_A)$ as a function of the strength of the interactions. Upon increasing the lattice size, $S_2$ approaches a step-like singular behavior. This singular behavior is a signature of bosonic criticality~\cite{frerot2016}. In Fig.~\ref{fig:5}(b), we rescale this quantity by the the dimension of the boundary $L$. As in Ref.~\cite{alba2013}, we observe a strict area-law dependence of the entanglement entropy in the Mott phase along with a logarithmic departure in the superfluid phase. In Fig.~\ref{fig:5}(c), we further extract area-law coefficient $a$ of Eq.~\eqref{eq:s2-scaling}. There, as predicted by Ref.~\cite{frerot2016}, one observes a pronounced cusp at the critical point, a signature of the contribution of the Higgs mode to the entanglement entropy.

\section{Conclusion\label{sec:5}}

In this paper, we introduced the neural backflow-Jastrow architecture. We physically motivated its relevance for the study of interacting lattice bosons and analytically proved that it encompasses the structure of various standard variational architectures while providing a natural generalization.
We benchmarked our Ansatz on the Bose-Hubbard problem and systematically assessed its variational accuracy, achieving a V-score as low as $2\times 10^{-4}$ on a $20{\times}20$ lattice.
Its great scalability allowed us to scale up VMC simulations to lattices with up to $20{\times}20$ sites on a single GPU with no local Hilbert-space truncation. Thanks to this, we performed a finite-size scaling analysis at zero temperature, showing remarkable agreement with the results obtained in QMC calculations by extrapolating finite-temperature results.
Furthermore, we were able to investigate the entanglement properties of the system. We observed the logarithmic correction of the entanglement entropy due to the spontaneous breaking of the $U(1)$ continuous symmetry and were able to extract its universal scaling prefactor, which we found to display a second-order singularity at the critical point.

This work clears the way for the simulation of many bosonic lattice systems which remain out of reach for other techniques. A prime example is the simulation of bosons in the presence of a synthetic magnetic field~\cite{weerda2024}. Indeed, such systems are intrinsically plagued with the sign problem, thereby ruling out QMC, and set in two spatial dimensions, which challenges matrix produc states, hitherto constrained to narrow-ladder geometries. Another interesting prospect is the extension of this architecture to driven-dissipative scenarios, where accurate neural representations of the density matrix are still unavailable. This would enlarge the scope of the NQS approach to the field of quantum optics.

\begin{acknowledgments}
We thank Federico Becca for his valuable early input and for conducting the GFMC simulations presented in this study. Our gratitude extends to Stephen R Clark for sharing data from Ref.~\cite{pei2024}. We also appreciate the helpful discussions with Markus Holzmann, Andreas M. Läuchli, David Clément, and Tommaso Roscilde. This work was supported by SEFRI through Grant No.\ MB22.00051 (NEQS - Neural Quantum Simulation).

All simulations were carried out with NetKet 3~\cite{vicentini2022}; the code used for this article can be found at Ref.~\cite{repository}.
\end{acknowledgments}

\bibliographystyle{unsrtnat}
\bibliography{references}

\onecolumn\newpage
\appendix

\section{Holon-doublon confinement\label{app:a}}

Close to the Mott-to-superfluid transition, the state of the system can be expressed as a perturbative expansion around the $U/J \sim +\infty$ limit. To leading order, this yields the following correction to the Mott state $\ket{\psi_\infty}$:
\begin{equation}
    \ket{\psi_U} - \ket{\psi_{\infty}} \propto \frac{J}{U}\sum_{\langle i,j\rangle}(\hat{a}_i^\dagger\hat{a}_j^{\mathstrut} + \hat{a}_j^\dagger\hat{a}_i^{\mathstrut})\ket{\psi_\infty} + O\bigl((J/U)^2\bigr).
\end{equation}
At unit filling this corresponds to the emergence of isolated holon-doublon bound pairs at any neighboring pair of sites. Similarly, higher-order contributions involve additional pairs, or pairs separated by a larger distance. Within the Mott phase, such an expansion converges and holon-doublon pairs must thus be spatially confined exponentially in order to stabilize the phase.

Such a confinement can be encoded in the variational wavefunction in the form of an attractive potential~\cite{capello2005}
\begin{equation}
    \ket{\psi'} = e^{-\sum_{ij}\hat{h}_i V_{d_{ij}} \hat{d}_j}\ket{\psi},
\end{equation}
where $V_d \leq 0$, for $0<d\leq R$, and $V_d = 0$ otherwise. Upon assuming any holon (doublon) is not surrounded by more than one doublon (holon), the many-body projector of Eq.~\eqref{eq:mbj} is a particular case of the above withrange $R=1$ and $V_1 = -g_{\mathrm{MB}}$. The $R>1$ case can be easily represented by a convolutional network. Indeed, upon neglecting occupations above $2$ bosons per site and assuming that pairs are distant from each other by more than $R$, one has
\begin{gather}
    - \sum_{ij}V_{d_{ij}}\bra{\vb*{n}}\hat{h}_i\hat{d}_j\ket{\vb*{n}} = \sum_{i}h_{i,1}^\pidx{2}(\vb*{n}),\\
    h_{i,1}^\pidx{2}(\vb*{n}) = \mathrm{ReLU}\Bigl(-V_0\sum_{\mu=1}^2 h_{i,\mu}^\pidx{1}(\vb*{n})-\sum_{d=1}^R V_d h_{i,2+d}^\pidx{1}(\vb*{n})\Bigr),\\
    h_{i,\mu\leq2}^\pidx{1}(\vb*{n}) = \mathrm{ReLU}\Bigl((-1)^{\mu}(n_i + 1)\Bigr),\\
    h_{i,\mu=2+d}^\pidx{1}(\vb*{n}) = \mathrm{ReLU}\Bigl(\sum_{\vb*{v}\in\mathcal{V}}\delta_{v,d}n_{T_{\vb*{v}}(i)}\Bigr),
\end{gather}
with $V_0 = \sum_{d>0}\lvert V_d\rvert$ and where input occupations are shifted by the unit density such that configurations $n=-1$ and $n=1$ correspond respectively to a hole and a doublon, such as in the main text. This bears the form of a two-layer convolutional network with parameters
\begin{gather}
    K^\pidx{1}_{\vb*{v},\mu\leq 2,1} = (-1)^\mu\delta_{v,0} \in \mathbb{R}^{\lvert\mathcal{V}\rvert\times 2\times 1},\quad b_\mu^\pidx{1} = (-1)^{\mu}\in\mathbb{R}^2,\nonumber\\
    K^\pidx{1}_{\vb*{v},\mu=2+d,1} = \delta_{v,d} \in \mathbb{R}^{\lvert\mathcal{V}\rvert\times R\times 1},\quad b_\mu^\pidx{1} = 0\in\mathbb{R}^R,\nonumber\\
    K^\pidx{2}_{\vb*{0},1,\mu'} = -V_\mu \in \mathbb{R}^{1\times1\times2},\quad b_1^\pidx{2} = 0\in\mathbb{R}.
\end{gather}

\section{Complexity of the neural backflow Jastrow Ansatz\label{app:b}}

We here assess the complexity of evaluating our variational wavefunction, as defined in Eq.~\eqref{eq:bfj}.

The translationally invariant two-body Jastrow operation has a complexity of $O(L^{2d})$ in $d$ spatial dimensions for the simplest algorithm. The neural backflow transformation of Eq.~\eqref{eq:backflow} has a complexity scaling as $O(L^d\times D \alpha W^d)$, with $D$ the depth of the network, $\alpha$ the number of convolutional features, $W$ the width of the filters, and $L$ the lattice lateral dimension.

The above estimation concerns the complexity of evaluating amplitudes of the Ansatz at fixed neural-network architecture. However, to keep a comparable relative error on the energy across system sizes, some hyperparameters of the variational wavefunction must carry a dependence in $L$. In any gapped phase, the system has quantum correlations with fixed correlation length irrespective of system size. In this regime, we expect the long-range correlations to be captured by the Jastrow, while short-range quantum correlations at a scale $\xi_c$ to be accounted for by a convolutional backflow transformation with a receptive field of size $\sim \xi_c^d$. In this regime, the complexity of the backflow transformation scales as $O(\alpha L^d \xi_c^d)$ for a shallow CNN and as $O(\alpha L^d \xi_c)$ for a deep one, and is thus always proportional to the system size. In a critical region, in contrast, the receptive field must be extensive $\sim L^d$, which incurs a complexity scaling of either $O(\alpha L^{2d})$ or $O(\alpha L^{2d+1})$ for a shallow or deep CNN, respectively.

Scaling the size of the system also has consequences in terms of the number of parameters, which ultimately impacts the complexity scaling of stochastic reconfiguration. In any regime, the Jastrow term has a number of parameters of $\sim dL/2$, whereas for the backflow transformation this scales as $O(\alpha \xi_c^d)$ for a shallow CNN and $O(\alpha \xi_c)$ for a deep one. It follows, that the backflow proposed in this work only involves a constant parameter overhead over Jastrow for gapped systems, and a parameter scaling similar to that of the bare Jastrow in critical regions for deep architectures.

From this analysis, it also appears that it is always better performance-wise to increase the depth of the backflow transformation as the system approaches the critical point rather than to increase the filter size.

\subsection{Metropolis-Hastings transition rule}

In all simulations, Markov-chain Monte Carlo (MCMC) was used in order to sample from the Born probability distribution $p_{\vb*{\theta}}(\vb*{n}) := \lvert\psi_{\vb*{\theta}}(\vb*{n})\rvert^2 / \lVert\psi_{\vb*{\theta}}\rVert^2$. At each MCMC step, an update from the current configuration $\vb*{n}$ into a new configuration $\vb*{n}'$ is proposed with probability given by the proposition distribution $g(\vb*{n}'\vert\vb*{n})$, and accepted with probability
\begin{equation}
    p_\mathrm{acc}(\vb*{n}\rightarrow\vb*{n}') = \min\left(1, \frac{p_{\vb*{\theta}}(\vb*{n}')}{p_{\vb*{\theta}}(\vb*{n})}\frac{g(\vb*{n}\vert\vb*{n}')}{g(\vb*{n}'\vert\vb*{n})}\right).
    \label{eq:metropolis}
\end{equation}

Our proposition distribution is induced by the first-quantization local transition rule $g^{\pidx{1}}(\vb{x}'\vert\vb{x}) = \prod_\mu^N g^{\pidx{1}}_\mathrm{loc}(\vb*{x}_\mu'\vert\vb*{x}_\mu)$ with 
\begin{equation}
    g_\mathrm{loc}^{\pidx{1}}(\vb*{x}_\mu'\vert\vb*{x}_\mu) \propto \mathds{1}\bigl[\lVert \vb*{x}_\mu - \vb*{x}_\mu'\rVert = a\bigr],
\end{equation}
where $\vb*{x}_\mu$ denotes the $\mu$th particle's position and $a$ is the lattice parameter. This proposal is clearly ergodic and corresponds to the Hamiltonian rule for a non-interacting system. It simply moves a particle picked at random into any of its neighboring sites. In second quantization, this can be achieved through the update kernel $g^{\pidx{2}}(\vb*{n}'\vert\vb*{n}) = \prod_{\langle i,j\rangle} g_\mathrm{loc}^{\pidx{2}}(n_i',n_j'\vert n_i, n_j)$ with
\begin{equation}
    g_\mathrm{loc}^{\pidx{2}}(n_i',n_j'\vert n_i, n_j) \propto n_i \delta_{n_i',n_i-1}\delta_{n_j',n_j+1}.
\end{equation}
This corresponds to choosing a site with a probability proportional to its occupation number and then transferring one of the occupations to any neighboring site. This kernel is not symmetric, hence, a correction is needed to enforce detail balance. This is accounted for in Eq.~\eqref{eq:metropolis} thanks to the following ratio of proposal densities:
\begin{equation}
    \frac{g_\mathrm{loc}^{\pidx{2}}(n_i,n_j\vert n_i', n_j')}{g_\mathrm{loc}^{\pidx{2}}(n_i',n_j'\vert n_i, n_j)} = \frac{n_j'}{n_i}.
\end{equation}

\section{Optimization procedure\label{app:c}}

In all simulations the learning rate was set to $\eta = \num{1e-3}$, the diagonal-shift regularization of the quantum geometric tensor was set to $\lambda = \num{5e-4}~(\num{1e-3})$ for the (neural backflow) Jastrow. $\num{8192}$ samples were used for most of the VMC calculations, and $\num{12288}$ for backflow transformations of depth $D=8$ for the largest lattice, while the number of samples was set to $\num{32768}$ for our simulations in Table~\ref{tab:1}.

\section{Worst-case variational figures of merit\label{app:d}}

Variational Monte Carlo proves to be the most demanding in the Mott phase at the vicinity of the critical point $U_c^+/J$. In Table~\ref{tab:2}, we compile the variational energies and corresponding variational scores in that challenging region ($U/J = 16.8$) as a function of system size and depth $D$ of the neural backflow transformation.

\begin{table*}[h]
\centering
\begin{tabular}{cclc}
Lattice size & Depth & \mc{$E_{\vb*{\theta}} / JL^{2}$} & V-score\\
\hline
\multirow{2}{*}{$8{\times}8$} & $0$ & $-0.452(2)$ & $\num{8.4e-2}$\\
 & $6$ & $-0.54783(6)$ & $\num{5.6e-5}$\\
\hline
\multirow{2}{*}{$12{\times}12$} & $0$ & $-0.444(1)$& $\num{7.9e-2}$\\
 & $6$ & $-0.54320(7)$& $\num{1.6e-4}$\\
\hline
\multirow{4}{*}{$16{\times}16$} & $0$ & $-0.446(1)$ & $\num{8.1e-2}$\\
 & $2$ & $-0.5375(2)$ & $\num{3.7e-3}$\\
 & $4$ & $-0.5411(1)$ & $\num{7.3e-4}$\\
 & $6$ & $-0.54191(7)$ & $\num{3.0e-4}$\\
\hline
\multirow{5}{*}{$20{\times}20$} & $0$ & $-0.4414(9)$ & $\num{7.8e-2}$\\
 & $2$ & $-0.5373(1)$ & $\num{3.1e-3}$\\
 & $4$ & $-0.54119(5)$ & $\num{6.8e-4}$\\
 & $6$ & $-0.54162(5)$ & $\num{3.1e-4}$\\
 & $8$ & $-0.54182(4)$ & $\num{1.8e-4}$\\
\end{tabular}
\caption{\label{tab:2}Variational figures of merit at $U/J = 16.8$. $\num{12288}$ samples were used in optimizing the networks of depth $D=6$ and $D=8$ for the largest system size, $\num{8192}$ otherwise.}
\end{table*}

\end{document}